\documentclass[]{iau}
\usepackage{natbib}
\usepackage{graphicx}

\title[50 Years of Candidate Pulsar Selection - What next?] 
{Fifty Years of Candidate Pulsar Selection \\ - What next?}

\author[R. J. Lyon]   
{R. J. Lyon$^1$}

\affiliation{$^1$School of Physics \& Astronomy, University of Manchester, \\ Oxford Road,
Manchester, UK, M13 9PL \\ email: {\tt robert.lyon@manchester.ac.uk} \\}

\pubyear{2017}
\volume{337}  
\jname{Pulsar Astrophysics: The Next Fifty Years}
\editors{P. Weltevrede, B.B.P. Perera, L. Levin Preston \& S. Sanidas, eds.}
\begin{document}

\maketitle

\begin{abstract}
For fifty years astronomers have been searching for pulsar signals in observational data. Throughout this time the process of choosing detections worthy of investigation, so called ‘candidate selection’, has been effective, yielding thousands of pulsar discoveries.  Yet in recent years technological advances have permitted the proliferation of pulsar-like candidates, straining our candidate selection capabilities, and ultimately reducing selection accuracy. To overcome such problems,  we now apply ‘intelligent’ machine learning tools.  Whilst these have achieved success, candidate volumes continue to increase, and our methods have to evolve to keep pace with the change. This talk considers how to meet this challenge as a community.
\end{abstract}

\firstsection 
\section{Introduction}
For fifty years we have diligently analysed the signal detections made through searches of the radio sky, in the hope of making new pulsar discoveries. Throughout that time, the process of selecting detections worthy of follow-up, so called `candidate selection', has been continually refined. What began during the earliest searches with analyses of pen chart records, has progressed to the application of statistical techniques capable of automatically identifying astrophysical signals with high accuracy. However, the proliferation of candidates exhibiting pulsar-like characteristics is placing increasing pressure on our selection capabilities. The accumulation of candidates has multiple causes, though is principally driven by technological improvements and changing science requirements. As technology continues to advance, this problem will worsen. Eventually survey/candidate data volumes will exceed our data storage capabilities due to cost \citep{Lyon:2016:bs}. In such scenarios it becomes important to prioritise those detections likely to yield a discovery for storage. Whilst increasing data rates introduce a time-critical facet to the selection problem - prioritisation must be done before a new batch of candidates arrive. To overcome these issues requires a shift from off-line to real-time data processing. This work considers how we meet this challenge as a community.
\section{Candidate Selection}
Candidate selection began as the search for periodic signal spikes in paper-based pen chart recordings. The success of this labour intensive process was entirely predicted on the skill of the individual performing the search, and their tenacity. Whilst this method yielded the first pulsar discovery \citep{Hewish:1968:jb}, it became out-dated with the advancement of computer technology and the digitization of the search process. Digitization enabled the application of filtering algorithms which initially achieved success. However at the same time, technological advancements were also increasing candidate yields \citep{Lyon:2016:bs}. This become a problem for \cite{Clifton:1986:tr}, as algorithmic filters alone were insufficient to select candidates effectively. In response many new methods were developed. From improved filtering approaches, to candidate ranking systems, and even graphical selection tools. Yet candidate volumes continued to increase, and the selection problem hardened. The community turned to sophisticated ML methods for help \citep[e.g.][]{Eatough:2010:uz}. These are able to automatically select candidates rapidly in an unbiased and reproducible way. The success achieved using ML has stimulated much work in this area. Thus ML-based selection now represents the state-of-the-art in this problem domain\footnote{Readers interested in a full review of the history of candidate selection, should refer to \cite{Lyon:2016:thesis} and \cite{Lyon:2016:bs}.}.
\section{Machine Learning}
ML is a branch of artificial intelligence (A.I.), concerned with replicating and improving upon the human ability to learn. Candidate selection employs tools from a specific area of ML known as pattern recognition - also known as statistical classification. The goal of classification is to automatically categorise data points as accurately as possible. Human beings are capable of undertaking complex classification tasks with ease, given appropriate training. This is due to our innate ability to learn via trial and error. ML algorithms learn in a similar way, however using statistical tools \citep[see][]{Bishop:2006:pr}.\newline

The aim of classification is to build functions that accurately map a set of input data points, to a set of class labels. For the candidate selection problem, this means mapping each candidate to its correct label (pulsar or non-pulsar). If $C=\lbrace X_{\rm 1}, \ldots , X_{\rm n} \rbrace $ represents a set of candidate data, then $X_{\rm i}$ is an individual candidate represented by variables known as \textit{features}. The features used must be chosen after careful consideration and analysis. It is desirable to use features that exhibit distributional differences between the pulsar and non-pulsar class. The features could include, for example, the folded signal-to-noise ratio (S/N) or the dispersion measure (DM). A label $y$ is associated with each candidate describing its true class. If the true class is unknown, then so too is the value of $y$.\newline

An ML function `learns' to separate candidates using data from a labelled vector called the training set. It contains the pairs $(X_{\rm 1},y_{\rm 1}),\ldots,(X_{\rm n},y_{\rm n})$. A classifier induces a mapping function between candidates and labels based on the training data. It does this by attempting to minimise the mapping errors made on the training examples. The trained function can then be used to label new unseen candidates in a `test' set. The test set may be comprised of an independent sample of examples used to test the trained classifier, or real world data that needs to be categorised. It is possible to deploy a classifier in both off-line and real-time processing scenarios, making ML suitable for solving our future selection challenges.
\section{Open Problems \& Recommendations}
There are a number of issues that reduce the selection accuracy of state-of-the-art ML methods. In the following sections these are described, and practical recommendations for overcoming them provided.
\subsection{Choosing an Approach}
It is impossible to know a priori which ML method will provide the best performance for a data set. This is known as the no-free lunch theorem \citep{Wolpert:1996:dh,Wolpert:2002:dh}. To proceed we must test as many algorithms as possible, to determine which is best for our data \citep{Duin:1996:rp,Salzberg:1997:sl,Janez:2006:dj}. When doing so we must remain approach agnostic, use an appropriate evaluation methodology \citep[which is difficult, see][]{Hand:2009:dj}, and not default to a popular or personally preferred method. A recent example of such an evaluation occurred in the medical domain \citep{Olson:2017:lc}. At present we do not generally choose algorithms in a principled way, and can improve in this area.
\subsection{Independent \& Identically Distributed Samples}
Learning algorithms only perform well, when the Independent and Identically Distributed (i.i.d) assumption holds. This fundamental assumption holds when the data used to train a classifier, is identically distributed to the data being classified\footnote{This can be explained via a simple analogy. A student preparing for an exam can only be expected to perform well, if the topic being revised matches the topic of the exam.}. Violations of the assumption lead to sub-optimal classifier performance. This has been demonstrated in an interactive resource supporting these proceedings \citep{Lyon:2017:sm}. It is important to note that in the presence of i.i.d violations, one cannot conclude that a learning algorithm is poor, or the wrong choice for a specific problem. Given the `right' information, the same algorithm could perform extremely well. To mitigate these issues we must use the correct data to train our algorithms (i.e. training and test data from the same source), remove sources of bias, and ensure robustness to over/under training \citep[see][]{Bishop:2006:pr}. 
\subsection{Distributional Change Over Time}
Changes in our data occurring over time, cause violations in the i.i.d assumption. There is evidence for such change in pulsar data \citep{Lyon:2016:thesis}, and interference is known to change over short and long time-scales due to human activity. Indeed it has been established that such changes severely affect the efficiency of pulsar search pipelines, and we should attempt to mitigate the effects of such variations when possible \citep{vanHeerden:2016:em}. When distributional change is likely, it is inappropriate to train static classification models to process incoming pulsar data. Rather it is better to use so called `on-line' classification systems. These are able to adapt to change over varying time-scales. So far only one such method has been developed \citep{Lyon:2016:bs}, and more work is needed.
\subsection{Open Data \& Standards}
Machine learning is no panacea. It can help us if our data is descriptive, otherwise its success is limited. To improve our methods we need to exploit our data effectively as a community. We require gold standard data sets from multiple telescopes and search pipelines. These can be used to study the nature of the candidate selection problem, and perhaps more crucially, train robust classification systems. If the data is correctly standardised, we can share data collected from many instruments, and evaluate our methods with greater rigour. A gold standard data repository would also enable principled feature evaluations. This would allow us to quantify the performance of currently deployed selection methods.
\section{The Future}
During the past fifty years we have witnessed three trends relevant to candidate selection. We have observed i) increasing survey data capture rates and total data volumes, ii) increasing candidate volumes, and iii) improvements in computational power tracking closely to Moore's Law \citep{Moore:1965:ge}. These three trends are likely to continue for the foreseeable future. To address increasing data volumes, we require appropriate data management tools, file formats, data standards, and well-defined metadata. These are essential if we are to successfully mine this information for new and exciting discoveries. To overcome increasing data capture rates, we will transition to real-time processing. Eventually we will have to trust automated systems to make candidate selection decisions for us in real-time. Such autonomy cannot be achieved without intelligent systems. Thus the adoption of machine learning methods is likely to accelerate in the coming years. This will be aided via the decreasing cost of accelerator cards, such as Graphics Processing Units (GPUs). GPU resources can be exploited to enable complex forms of machine learning, previously impractical to implement due to computational cost. Those undertaking candidate selection will have to become familiar with new hardware and software infrastructures, to enable these resources to be properly exploited. Future astronomers will most likely need to be capable physicists, programmers, and machine learning practitioners, in order to mitigate the candidate selection problems of the future. This is a far cry from where it all began - a single astronomer with a paper-based record.


\begin{thebibliography}{}

\bibitem[Bishop (2006)]{Bishop:2006:pr} {Bishop} C.~M., 2006, Pattern Recognition and Machine Learning, Springer.

\bibitem[Clifton \& Lyne (1986)]{Clifton:1986:tr} {Clifton} T.~R. and {Lyne} A.~G., 1986, \textit{Nature}, vol.320, pp.43--45.

\bibitem[Duin (1996)]{Duin:1996:rp} {Duin} R.~P., 1996, Pattern Recognition Letters, vol.17(5), pp.529--536, DOI:10.1016/0167-8655(95)00113-1.

\bibitem[Eatough \etal\ (2010)]{Eatough:2010:uz} {Eatough} R.~P., {Molkenthin} N., {Kramer} M., {Noutsos} A., {Keith} M.~J., {Stappers} B.~W., {Lyne} A.~G., 2010, \textit{MNRAS}, 407, 2443.

\bibitem[Fawcett (2006)]{Fawcett:2006:tf} {Fawcett} T., 2006, Pattern Recognition Letters, vol.27(8), pp.861--874, DOI:10.1016/j.patrec.2005.10.010.

\bibitem[Hand (2009)]{Hand:2009:dj} {Hand} D.~J., 2009, Machine Learning, vol.77(1), pp.103--123, DOI: 10.1007/s10994-009-5119-5.

\bibitem[Hewish \etal\ (1968)]{Hewish:1968:jb} {Hewish} A., {Bell} S.~J., {Pilkington} J.~D.~H., {Scott} P.~F., {Collins} R.~A., 1968, \textit{Nature}, vol.217(5130), pp.709--713.

\bibitem[Janez (2006)]{Janez:2006:dj} {Janez} D., 2006, ``Statistical comparisons of classifiers over multiple data sets", Journal of Machine learning research, vol.7.

\bibitem[Lyon \etal\ (2016)]{Lyon:2016:bs} {Lyon} R.~J., {Stappers} B.~W., {Cooper} S., {Brooke} J.~M., {Knowles} J.~D., 2016, \textit{MNRAS}, vol.459(1):1104-1123, DOI: 10.1093/mnras/stw656.

\bibitem[Lyon (2016)]{Lyon:2016:thesis} {Lyon} R.~J., 2016, ``Why are Pulsars Hard to Find?", PhD Thesis, University of Manchester.

\bibitem[Lyon (2017)]{Lyon:2017:sm} {Lyon} R.~J., 2017, ``Supporting Material: Fifty Years of Candidate Pulsar Selection - What next?", DOI: 10.5281/zenodo.883844.

\bibitem[Moore (1965)]{Moore:1965:ge} {Moore} G.~E., 1965, ``Cramming more components onto integrated circuits", Electronics, vol.38(8).

\bibitem[Olson \etal\ (2017)]{Olson:2017:lc} {Olson} R.~S., {La Cava} W., {Mustahsan} Z., {Varik} A., {Moore} J.~H., 2017, ArXiv e-prints, q-bio.QM, arXiv:1708.05070.

\bibitem[Salzberg (1997)]{Salzberg:1997:sl} {Salzberg} S.~L., 1997, Data Mining and Knowledge Discovery, vol. 1(3). DOI: 10.1023/A:1009752403260.

\bibitem[van Heerden (2016)]{vanHeerden:2016:em} {van Heerden} E., {Karastergiou} A. and {Roberts} S.~J., 2016, \textit{MNRAS}, vol. 467(2), pp.1661--1677,  DOI: 10.1093/mnras/stw3068.

\bibitem[Wolpert (1996)]{Wolpert:1996:dh} {Wolpert} D.~H., 1996, Neural computation, vol. 8(7), pp.1341--1390.

\bibitem[Wolpert (2002)]{Wolpert:2002:dh} {Wolpert} D.~H., 2002, ``The Supervised Learning No-Free-Lunch Theorems", In Soft Computing and Industry, pp.25-–42. Springer London.
  


\end{thebibliography}
\end{document}